\documentclass[aps,pra,twocolumn,amsmath,showpacs]{revtex4}
\usepackage{bm}
\usepackage{graphicx}
\begin{document}
\setlength{\arraycolsep}{2pt}
\title{Linear optical schemes to demonstrate {\it genuine} multipartite entanglement for single-particle $W$ states}
\author{Hyunchul Nha$^*$}
\affiliation{Department of Physics, Texas A \& M University at Qatar, Doha, Qatar} 
\date{\today}
%\maketitle
\begin{abstract}
We consider the method of entanglement witness operator to verify {\it genuine} multipartite entanglement for single-particle 
$W$ states involving $N$ parties. In particular, linear optical schemes using photo detectors and beam splitters are proposed to 
implement two different types of witness operator in experiment. 
The first scheme that requires only a {\it single} measurement setting 
is shown to detect genuine multipartite entanglement 
for the overall efficiency beyond $1-1/N$. 
On the other hand, the second scheme with $N+1$ measurement settings achieves success at a significantly lowered efficiency than 
$1-1/N$. 
\end{abstract}
\maketitle
\email{hyunchul.nha@qatar.tamu.edu}

\narrowtext
\section{Introduction}

Entanglement, an element of central importance in quantum information processing, is still far from our complete understanding 
despite a great number of efforts for the past decades. 
Earlier works on bipartite entanglement \cite{bi} have been naturally extended to multipartite systems, and the multipartite entanglement 
has been identified as a resource to implement quantum information processing to larger extent, 
such as error correction \cite{Shor}, secret sharing \cite{Cleve}, and one-way quantum computing \cite{onew}, 
to name a few. Nevertheless, multipartite entanglement has been far less understood, and its generation/verification 
in experiment seems more demanding than the bipartite ones \cite{multi}. 

One of the important issues regarding many-body quantum systems is to verify {\it genuine} multipartite entanglement 
among parties in one way or another. Genuine multipartite entanglement is distinguished from other types of entanglement 
by the participation of {\it all} parties in quantum correlations, and it is particularly distinct from biseparable entanglement: 
Consider an $N$-party quantum system represented by the state of form $|\Psi\rangle\otimes|\Phi\rangle$, 
where $|\Psi\rangle$ belongs to the Hilbert space $H^{\otimes m}$ ($m$ parties) and  $|\Phi\rangle$ to $H^{\otimes N-m}$ 
($N-m$ parties). Although this system can show quantum correlations to some degree among $m$ parties and $N-m$ parties, 
respectively, it is far from true $N$-party entanglement. In general, a mixed state $\rho$ is called biseparable 
if it is a mixture of pure biseparable states, with different bipartitions allowed among component systems. 
Otherwise, the state is genuinely multipartite entangled. 
One example of biseparable states is the tripartite system represented in the number-state basis as
\begin{eqnarray}
\rho_{123}=&&\frac{1}{3}|B\rangle\langle B|_{12}\otimes|0\rangle\langle0|_3+\frac{1}{3}|B\rangle\langle 
B|_{23}\otimes|0\rangle\langle0|_1\nonumber\\
&&+\frac{1}{3}|B\rangle\langle B|_{31}\otimes|0\rangle\langle0|_2,
\end{eqnarray}
where $|B\rangle$ is the Bell state of two parties and $|0\rangle$ the vacuum state of the third \cite{Guhne0}.

Given an $N$-body system, it is an important, but nontrivial, task to determine whether the system possesses genuine 
multipartite entanglement. One possible approach to this problem is to set up the entanglement witness operator 
${\cal W}$ \cite{Acin, Lewenstein,Guhne1,Sanpera,Toth} in such a way that the ensemble average ${\rm tr}\{\rho {\cal W}\}$ 
becomes positive or zero for all biseparable states \cite{Bell}. If it takes a negative value, it thus becomes a clear signature 
of genuine multipartite entanglement. In this paper, we consider the detection of genuine multipartite entanglement 
for single-particle $W$ states based on the entanglement witness \cite{Eibl}. Particularly, we are interested 
in the implementation of the idea in an optical experiment involving beam splitters and photo detectors. 
Note that the same problem was previously investigated, but in a limited context \cite{nha0}. 
Specifically, Nha and Kim showed that the pairwise entanglement between {\it arbitrarily} chosen two modes can be detected 
regardless of the photo detector efficiency \cite{nha0} using the entanglement conditions derived for continuous variables 
\cite{nha1,nha2,Hillery,Agarwal}. Of course, it may indicate the multipartite entanglement structure of the $W$ states 
to some extent \cite{Dur}, however, it is not a rigorous proof of {\it genuine} multipartite entanglement \cite{Guhne0}. 
A counter-example is the biseparable state in Eq.~(1), for which {\it any} two modes, when the third mode is traced out, 
have {\it nonzero} pairwise entanglement.

In this paper, we consider two types of witness operator for $W$ states and propose how to implement those 
in practice for single-photon $W$ states \cite{Lombardi}. 
The first scheme, which requires only a {\it single} measurement setting without the need of a full state tomography, 
is shown to succeed in verifying genuine multipartite entanglement for the overall efficiency over $1-1/N$, 
where $N$ is the total number of modes. On the other hand, the second scheme with $N+1$ measurement settings 
is shown to achieve success even at a lower efficiency than $1-1/N$. 

%In addition, we discuss the optimization of measurement scheme for the case of non-symmetric $W$ states. 
The proposed scheme in this paper is derived from the observation 
that the quantum fidelity of a given state $\rho$ with respect to an entangled state $|E\rangle=U|P\rangle$ is the same as the fidelity of the transformed state $\rho'\equiv U^{\dag}\rho U$ with the product state $|P\rangle$. Here, the operation $U$ is the entangling action on the product state $|P\rangle$ to obtain the entangled state $|E\rangle$. 
Therefore, the fidelity measurement is alternatively achieved after the inverse unitary operation $U^\dag$ is performed on the state $\rho$.   
Note that a similar idea based on the time-reversed operations to demonstrate many-body quantum coherences has been well known in the nuclear magnetic resonance (NMR) community \cite{Knill}. In particular, Lee and Khitrin recently used the time-reversed sequences of entangling operations, i.e. disentangling operation, to verify the 12-spin "Schr{\"o}dinger-cat" state in \cite{Lee}.

This paper is organized as follows. In Sec.~II, we briefly introduce the entanglement witness operator and specify it 
for the class of $W$ states. In Sec.~III, a general idea of measuring the quantum fidelity, which is essential 
for the implementation of witness operator, is presented and applied to the case of single-photon $W$ states. 
In Sec.~IV, the proposed scheme involving only a {\it single} measurement setting is analyzed with some experimental inefficiencies 
incorporated, and in particular, the optimization of the detection scheme for the case of asymmetrical $W$ states is discussed. 
In Sec.~V, an improved scheme with $N+1$ measurement settings is presented along with a modified witness operator, 
and the critical efficiency for successful entanglement detection is shown to be significantly lowered than $1-1/N$.
Finally, the main results of this paper are summarized in Sec.~VI.

\section{Entanglement Witness Operators for $W$-class states}
In this section, we briefly introduce the entanglement witness operator to detect genuine multipartite entanglement 
and specify it for the class of $N$-partite $W$ states. Suppose that one has produced a certain multipartite state $\rho$, 
presumably entangled and most likely mixed due to experimental imperfections, close to a {\it target} pure entangled state 
$|\Psi\rangle$. An witness operator ${\cal W}$ can then be constructed in a form 
\begin{eqnarray}
{\cal W}=\alpha I-|\Psi\rangle\langle\Psi|,
\label{eqn:witness}
\end{eqnarray}
where the constant $\alpha$ is taken as the maximum possible overlap of a pure biseparable state $|\phi\rangle$ 
with the genuine multipartite entangled state $|\Psi\rangle$.
Namely, 
\begin{eqnarray}
\alpha={\rm max}_{|\phi\rangle\in {\rm B}}|\langle\phi|\Psi\rangle|^2,
\label{eqn:alpha}
\end{eqnarray}
where B represents the set of biseparable states \cite{Sanpera}. It is now straightforward to see ${\rm tr}\{\rho {\cal W}\}\ge0$ 
for every biseparable state $\rho=\rho_{\rm BS}$ so that only genuine multipartite entangled states can take a negative value 
over the witness operator $\cal W$. This is generally true for mixed states due to the linearity of the witness operator 
and the convexity of separable states. 

Although it seems very demanding to find out the value of $\alpha$ for a given state $|\Psi\rangle$ through optimization 
in Eq.~(\ref{eqn:alpha}), Bourennane {\it et al.} identified the constant $\alpha$ with the maximum Schmidt coefficient 
of $|\Psi\rangle$ with respect to all bipartite settings \cite{Sanpera}. More concretely, in a fixed bipartition of all parties, 
say ${\cal B}^{(k)}$ ($k$-th bipartition), one may choose an orthonormal product basis $|ij\rangle$, 
where $|i\rangle$ and $|j\rangle$ belong to the Hilbert spaces of two parties, respectively. 
Then, the target state $|\Psi\rangle$ is represented in the same basis as $|\Psi\rangle=\Sigma_{ij}C_{ij}^{(k)}|ij\rangle$ 
and the maximum singular value $\lambda_{\rm max}^{(k)}$ of the matrix $C_{ij}^{(k)}$ can be evaluated for the $k$-th bipartition. 
The same steps must be taken to obtain $\lambda_{\rm max}$ for every possible bipartition and $\alpha$ corresponds to the maximum 
among those $\lambda_{\rm max}^{(k)}$.  

Let us now consider the class of $N$-partite $W$ states, which is represented by a form
\begin{eqnarray}
|W\rangle=c_1|1,0,\cdots,0\rangle&+&c_2|0,1,\cdots,0\rangle\nonumber\\+\cdots&+&c_N|0,0,\cdots,1\rangle.
\label{eqn:w}
\end{eqnarray} 
Then, the constant $\alpha$ in Eq.~(\ref{eqn:alpha}) is evaluated as
\begin{eqnarray}
\alpha=1-{\rm min}\{|c_i|^2\},
\label{eqn:alpha2}
\end{eqnarray} 
where ${\rm min}\{|c_i|^2\}$ is the minimum among all $|c_i|^2$.
Given the number of parties $N$, the constant $\alpha$ takes the smallest value $1-1/N$ for the class of symmetric $W$ states 
($c_i=1/\sqrt{N}$ for all $i$), which may be thus less demanding to detect than asymmetric $W$ states.

\section{Fidelity Measurement}
In this section, we first present a general idea to implement the witness operator in experiment and then apply it to the class of 
single-photon $W$ states. Once the constant $\alpha$ is identified in Eq.~(\ref{eqn:alpha}), the remaining task is to measure the 
fidelity between the reference state $|\Psi\rangle$ and the state $\rho$ in question, namely, 
$\langle\Psi|\rho|\Psi\rangle$ \cite{Jozsa}, as the quantum average of the witness operator becomes 
${\rm tr}\{\rho {\cal W}\}=\alpha-\langle\Psi|\rho|\Psi\rangle$. 

For the case of continuous variables, Kim {\it et al.} have proposed an experimental scheme to measure fidelity by mixing 
the two fields at a beam splitter and measuring the Wigner function of the output \cite{Kim}. In particular, they considered 
homodyne detection to measure the fidelity of two Gaussian field states \cite{nha3}. However, this approach, which needs 
the preparation of two states to compare, may not be suited to the case of witness operator:  
The reference state $|\Psi\rangle$ in the witness operator is only a target and one shall have produced instead a mixed state 
$\rho$ in reality. In other words, one has no reference state to compare with the subject $\rho$. 
In this paper, another method to experimentally evaluate the fidelity, which does not rely on the preparation of reference state 
and proceeds only with the state $\rho$, is proposed as follows.

In a number of cases, a many-body entangled state $|\Psi\rangle$ can be generated from an initial product state 
$|\phi_1\rangle|\phi_2\rangle\cdots|\phi_N\rangle$ which subsequently undergoes some entangling unitary operations collectively 
represented by $U$. Namely,
\begin{eqnarray}
|\Psi\rangle=U|\phi\rangle\equiv U|\phi_1\rangle|\phi_2\rangle\cdots|\phi_N\rangle.
\end{eqnarray}
Then, the fidelity of our concern can be expressed as 
\begin{eqnarray}
\langle\Psi|\rho|\Psi\rangle=\langle\phi|\rho'|\phi\rangle,
\end{eqnarray}
where
\begin{eqnarray}
\rho'\equiv U^{\dag}\rho U.
\end{eqnarray}
Consider the case that the initial product state is a collection of number states, i.e., 
$|\phi\rangle=|\phi_1\rangle|\phi_2\rangle\cdots|\phi_N\rangle=|n_1\rangle|n_2\rangle\cdots|n_N\rangle$. 
Given the multimode state $\rho$, the fidelity is then  reduced to the photon counting distribution of the transformed state 
$\rho'\equiv U^{\dag}\rho U$. 

More precisely, the state $\rho$ is first subjected to the {\it inverse} unitary operation $U^{\dag}$ to create a new state 
$\rho'=U^{\dag}\rho U$. Next, one measures the joint probability $P_{n_1n_2\cdots n_N}$ that the photo detector at mode $i$ counts 
$n_i$ photons for the state $\rho'$, that is,
\begin{eqnarray}
P_{n_1n_2\cdots n_N}=\langle n_1|\langle n_2|\cdots\langle n_N|\rho'|n_1\rangle|n_2\rangle\cdots|n_N\rangle.
\end{eqnarray}
This probability corresponds to the fidelity of our concern.

Let us apply the above idea to the case of single-photon $W$ states. 
An arbitrary $N$-partite $W$-state of the form in Eq.~(\ref{eqn:w}) can be prepared by injecting a single photon 
into an array of beam splitters as shown in Fig.~1 \cite{nha0}. Namely, $|W\rangle=U|1,0,\cdots,0\rangle$, 
where the unitary operator $U$ is a series of beam splitter actions, $U=B_{\{N-1N\}}\cdots B_{\{12\}}$. 
The beam splitter operator $B_{\{jj+1\}}$ transforms two adjacent modes \{$a_j$, $a_{j+1}$\} into \{$a'_j$, $a'_{j+1}$\} as
\begin{eqnarray}
\begin{pmatrix}&a'_j\\&a'_{j+1}
\end{pmatrix}
=\begin{pmatrix}
&\sin\theta_j&-\cos\theta_j\\&\cos\theta_j&\sin\theta_j
\end{pmatrix}
\begin{pmatrix}&a_j\\&a_{j+1}
\end{pmatrix},
\end{eqnarray}
where $\cos\theta_i$ ($\sin\theta_i$) denote the transmissivity (reflectivity) of the beam splitter \cite{Campos}. 
The coefficients $c_i$ in Eq.~(\ref{eqn:w}) are then given by 
\begin{eqnarray}
c_1&=&\sin\theta_1,\nonumber\\ 
c_2&=&\cos\theta_1\sin\theta_2,\nonumber\\ 
&&\vdots\nonumber\\
c_j&=&\left[\Pi_{i=1}^{j-1}\cos\theta_i\right]\sin\theta_j,\nonumber\\ 
&&\vdots\nonumber\\
c_N&=&\left[\Pi_{i=1}^{N-1}\cos\theta_i\right],
\label{eqn:coeff}
\end{eqnarray}
at the output.  
If a phase shift is additionally carried out with the amount $\phi_j$  
at the $j$-th output mode, the coefficients become $\tilde{c}_j=c_je^{-i\phi_j}$ ($j=1,\cdots,N$).

\begin{figure}
\includegraphics[width=2.5in,keepaspectratio=true]{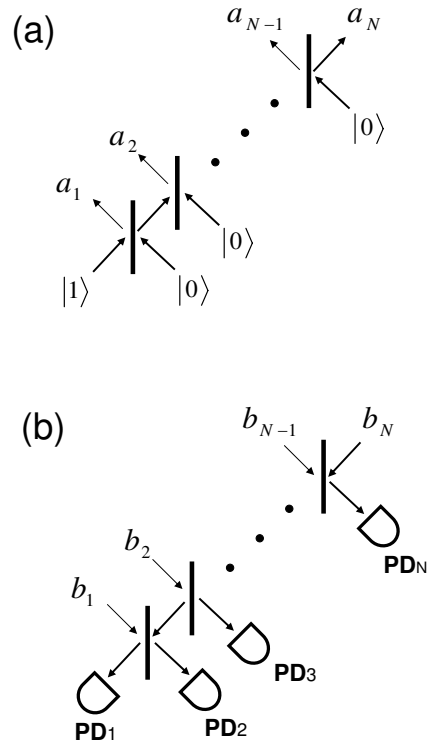}
\caption{(a) Experimental scheme for generating an arbitrary single-photon $W$-state in Eq.~(\ref{eqn:w}). 
A single-photon at one input is injected into an array of beam splitters with the transmissivity of the $j$-th beam splitter 
given by $\cos\theta_j$. Then, the coefficients $c_i$ in Eq.~(\ref{eqn:w}) are given as those in Eq.~(\ref{eqn:coeff}). 
(b) Experimental scheme for detecting {\it genuine} multipartite entanglement based on the $W$-class witness operator. 
A given $N$-partite state $\rho$, with each mode represented by $b_i$, is inversely injected as shown and each photo detector 
(PD) counts the photons at the output. See main text.}
\label{fig:lodi1}
\end{figure}

The above generation scheme suggests that the fidelity $\langle W|\rho|W\rangle=\langle 1,0,\cdots,0|U^{\dag}\rho 
U|1,0,\cdots,0\rangle$ can be measured by injecting a given state $\rho$ to the series of beam splitter in {\it reverse} 
order to produce the state $U^{\dag}\rho U$ and then measuring the counting probability $P_{10\cdots0}$ that only the first 
detector counts one photon and the others no photons.

\section{Detection of mixed $W$ states in an optimized setting}
In this section, we analyze the detection of genuine multipartite entanglement of the $W$-state in Eq.~(\ref{eqn:w}) 
based on the scheme outlined in Sec.~III. 
In particular, we take into account the experimental imperfections such as non-perfect single photon source and inefficient 
photo detectors. 

Suppose that one creates a single photon source with a success probability $p_S$ represented by a mixed state 
$\rho_{\rm single}=p_S|1\rangle\langle1|+(1-p_S)|0\rangle\langle0|$ \cite{Lvovsky}, which is injected to the beam-splitter array 
in Fig.~1. Then, the output state generated becomes 
\begin{eqnarray}
\rho=p_S|W\rangle\langle W|+(1-p_S)|0\cdots0\rangle\langle0\cdots0|,  
\label{eqn:actual}
\end{eqnarray}
and one wishes to detect the genuine multipartite entanglement of this mixed state.

When this output entangled state is subjected to the inverse array of beam splitters in the {\it same} configuration 
as the one used in its generation (Fig.~1), the counting probability $P_{10\cdots0}$ becomes $\eta p_S$, where $\eta$ is 
the efficiency of the photo detectors. Now, the detection of genuine multipartite entanglement turns out to be a success 
under the condition
\begin{eqnarray}
\alpha<\eta p_S,
\label{eqn:sucess}
\end{eqnarray}  
where the constant $\alpha$ is specified in Eq.~(\ref{eqn:alpha2}). In particular, for the symmetric $W$ states, 
the above condition reads as 
\begin{eqnarray}
1-\frac{1}{N}<\eta p_S,
\label{eqn:sucess1}
\end{eqnarray}  
in which the requirement of the overall efficiency, $\eta p_S$, becomes the least demanding. 

In the above analysis, we have considered the {\it same} configuration of the beam splitters to generate and detect 
the multipartite $W$ states. That is, the pure $W$ state component of the actual state $\rho$ in Eq.~(\ref{eqn:actual}) is 
the same as the $W$ state used as the reference in the witness operator $\cal W$. On the other hand, for the case of the states 
$\rho$ asymmetrical under permutations, one can try to optimize the detection scheme by adjusting the reference state as follows. 
Given the actual state $\rho$ in Eq.~(\ref{eqn:actual}), which is a mixture of the target state $|W\rangle$ and the vacuum state, 
one may choose a different witness operator ${\cal W'}$ with other $W$ state $|W'\rangle$ in Eq.~(\ref{eqn:witness}) as reference. 
Then the condition for the overall efficiency to detect genuine multipartite entanglement becomes
\begin{eqnarray}
\frac{\alpha'}{|\langle W|W'\rangle|^2}<\eta p_S,
\label{eqn:opt}
\end{eqnarray}
where $\alpha'$ is the constant for the state $|W'\rangle$ in Eq.~(\ref{eqn:alpha}).
Now, given the target state $|W\rangle$, it is desired to minimize the value $\frac{\alpha'}{|\langle W|W'\rangle|^2}$ by finding 
out the optimal reference state $|W'\rangle$.

As an example, let us consider an asymmetric tripartite $W$ state, $|W_a\rangle=\frac{1}{2}|1,0,0\rangle
+\frac{1}{2}|0,1,0\rangle+\frac{1}{\sqrt{2}}|0,0,1\rangle$. The constant $\alpha$ for this state is 3/4=0.75 from 
Eq.~(\ref{eqn:alpha2}), so if one uses the state $|W_a\rangle$ itself as the reference in the witness operator, 
the requirement becomes $0.75<\eta p_S$. On the other hand, by minimizing the value of $\frac{\alpha'}{|\langle W_a|W'\rangle|^2}$ 
in Eq.~(\ref{eqn:opt}), one finds $12-8\sqrt{2}\approx0.686<\eta p_S$ with the symmetric $W$ state 
$|W_s\rangle=\frac{1}{\sqrt{3}}|1,0,0\rangle+\frac{1}{\sqrt{3}}|0,1,0\rangle+\frac{1}{\sqrt{3}}|0,0,1\rangle$ as the reference, 
$|W'\rangle=|W_s\rangle$.

For a fixed number of parties, $N$, however, one can readily show that the value of $\frac{\alpha'}{|\langle W|W'\rangle|^2}$ 
is tightly bounded as
\begin{eqnarray}
\frac{\alpha'}{|\langle W|W'\rangle|^2}\ge1-\frac{1}{N}.
\label{eqn:bound}
\end{eqnarray}
The equality in Eq.~(\ref{eqn:bound}) holds for the symmetric $W$ states, 
$|W\rangle=|W'\rangle=\frac{1}{\sqrt{N}}|1,0,\cdots,0\rangle+\frac{1}{\sqrt{N}}|0,1,\cdots,0\rangle+\cdots
+\frac{1}{\sqrt{N}}|0,\cdots,0,1\rangle$. Therefore, in the present scheme, the overall efficiency $\eta p_S$ must be greater 
than $1-1/N$ to detect genuine $N$-partite entanglement and the choice of symmetric $W$ states is regarded as best 
from a practical point of view.

As a final remark, let us address the problem of local phase shifts that may occur in the multipartite state under test. 
Suppose that the prepared state is in a form 
\begin{eqnarray}
|W'\rangle=\frac{1}{\sqrt{N}}&&\left(e^{-i\phi_1}|1,0,\cdots,0\rangle+e^{-i\phi_2}|0,1,\cdots,0\rangle\right.\nonumber\\&&+\cdots
+e^{-i\phi_N}|0,\cdots,0,1\rangle\left.\right), 
\end{eqnarray}
where $\phi_i$ is the local phase shift of $i$-th mode.
If these phases are completely unknown, we cannot figure out a {\it single} experiment setup appropriately, 
as our scheme relies on the inverse operation to disentangle a given state. Even though the given state is genuinely 
multipartite-entangled, the measured fidelity $|\langle W|W'\rangle|$ could be zero in the worst case. 
To resolve this issue, if it arises, one must place a phase-shifter at each mode in the detection scheme of Fig.1~(b) before each mode enters the array of beam-splitters. 
By covering the whole range of phase-shift at each mode, one can maximize the fidelity, 
which will lead to a success in entanglement detection.
As the phase shifts are locally performed, this does not affect our judgment on the entanglement structure of the given state.

\section{Improved experimental scheme with $N+1$ measurement settings}

In this section, we discuss a possible improvement of the experimental condition by considering a modified witness operator. 
In Ref.~\cite{Blatt}, H{\"a}ffner {\it et al.} introduced the witness of the form 
\begin{eqnarray}
{\cal W'}=\alpha I_2-Q,
\label{eqn:witness_mod}
\end{eqnarray}
where 
\begin{eqnarray}
Q\equiv|W_N\rangle\langle W_N|-\beta\sum_{i=1}^N|BS_i\rangle\langle BS_i|.
\label{eqn:q}
\end{eqnarray}
Note that the identity operator $I_2$ in Eq.~(\ref{eqn:witness_mod}) refers to the Hilbert subspace 
corresponding to a total of two quanta.
In Eq.~(\ref{eqn:q}), the state $|BS_i\rangle=|0_i\rangle|W_{N-1}\rangle$ $(i=1,\cdots,N)$ is biseparable 
where the $i$-th mode is in the vacuum state $|0\rangle$ and the other $N-1$ modes in the $W$ state. 
The modified witness ${\cal W'}$ differs from the one in Eq.~(\ref{eqn:witness}) by subtracting from the $W$-state $|W_N\rangle$ 
the biseparable states $|BS_i\rangle$, which all give the maximum overlap, $1-1/N$, with $|W_N\rangle$. 
The constant $\alpha$ in Eq.~(\ref{eqn:witness_mod}) can be obtained by maximizing the expectation value of $Q$ 
over biseparable states 
as \begin{eqnarray}
\alpha={\rm max}_{|\phi\rangle\in {\rm B}}\langle\phi|Q|\phi\rangle.
\label{eqn:alpha_mod}
\end{eqnarray}
The optimal biseparable state $|\phi\rangle=|a\rangle|b\rangle$ for the maximum was identified as a form of 
$|a\rangle=\cos\theta_1|0,\cdots,0\rangle_k+\sin\theta_1|W_k\rangle$ and 
$|b\rangle=\cos\theta_2|0,\cdots,0\rangle_{N-k}+\sin\theta_2 |W_{N-k}\rangle$, 
respectively, for a fixed bipartition of $\{k,N-k\}$ modes \cite{Blatt}. Therefore, $\alpha$ can be numerically evaluated 
by optimizing the expectation value over the parameters $\theta_1,\theta_2$ and $k$ for a fixed value of $\beta$. 

To implement the modified witness operator ${\cal W'}$ in experiment, 
given a certain state $\rho$, we need to measure ${\rm tr}\{\rho I_2\}$ and 
$\langle BS_i|\rho|BS_i\rangle$ ($i=1,\cdots,N$) in addition to $\langle W_N|\rho|W_N\rangle$. 
The ensemble average ${\rm tr}\{\rho I_2\}$ is simply the probability that the total quanta is at most two, which can be measured in the same experimental setup as Fig.1 (b) with no extra efforts. In fact, it does not make a significant difference in our case if we take $I$ (entire identity operator), instead of $I_2$, for the witness ${\cal W'}$ in Eq.~(\ref{eqn:witness_mod}). On the other hand, the fidelity $\langle BS_i|\rho|BS_i\rangle$ can be measured in a similar setup to the one in Fig.1 (b) 
by injecting $i$-th mode directly to the photo detector and the other modes to the inverse beam-splitter array of the $N-1$ mode $W$-state 
$|W_{N-1}\rangle$. Therefore, a total of $N+1$ measurement settings are required to implement the witness operator 
${\cal W'}$. 

\begin{figure}
\includegraphics[width=2.5in,keepaspectratio=true]{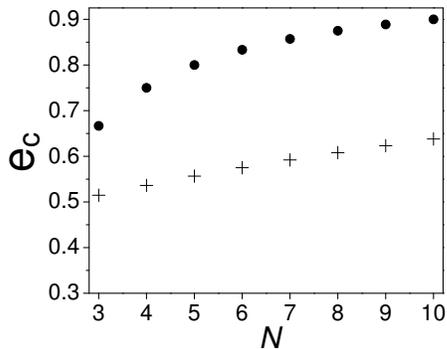}
\caption{Critical efficiency $e_c\equiv\frac{\alpha}{1-(N-1)\beta}$ for the modified scheme 
of Sec.~V involving $N+1$ measurement settings (cross), in comparison with the critical value $e_c=1-1/N$ 
for the single measurement scheme of Sec.~IV (circle), 
as a function of the number $N$ of modes. The genuine multipartite entanglement is successfully detected for the efficiency $\eta p_S>e_c$ in each case. 
The parameter $\beta$ in the witness ${\cal W'}$ of Eq.~(\ref{eqn:witness_mod}) 
was chosen as $\beta(N-1)=1-10^{-3}$.}
\label{fig:lodi2}
\end{figure}

Let us now consider the detection of symmetric $W$ states using the witness ${\cal W'}$.
With the same experimental imperfections characterized by $\eta$ and $p_S$ as in Sec.~IV, we obtain 
${\rm tr}\{\rho_W {\cal W'}\}=\alpha-\eta p_S[1-(N-1)\beta]$. 
Therefore, the detection of genuine multipartite entanglement is a success for the efficiency 
$\eta p_S>\frac{\alpha}{1-(N-1)\beta}$. 
As the constant $\alpha$ is given by the maximization procedure outlined above for a fixed $\beta$, one can finally obtain 
the improved experimental condition by minimizing the fraction $e_c\equiv\frac{\alpha}{1-(N-1)\beta}$ over the parameter $\beta$. 
We numerically checked that this fraction decreases as $(N-1)\beta\rightarrow1$ for a given number $N$. 
The result is plotted as a function of $N$ with the value of $(N-1)\beta=1-10^{-3}$ in Fig.~2. 
As clearly seen, the previous critical value of $1-1/N$ is 
significantly reduced in the modified scheme. For instance, when $N=3$, 
the improved condition becomes $e_c\approx0.515<\eta p_S$ for $2\beta=1-10^{-3}$ and $\alpha\approx5.15\times10^{-4}$.

\section{Summary} 

In this paper, the approach based on the witness operator was considered to detect {\it genuine} multipartite entanglement 
for the single-photon $N$-partite $W$ states. In particular, the experimental schemes using the photo detectors and 
the beam splitters were proposed to implement two different types of witness operator.  
In the first scheme involving only a single measurement setting, the threshold overall efficiency was found to be $1-1/N$ 
for the symmetrical mixed $W$ states, which therefore becomes increasingly hard with the number of parties $N$. 
On the other hand, in the second scheme involving a total of $N+1$ measurement settings, 
an improved condition was obtained with a significantly lowered critical efficiency. 

%It was also discussed that one can optimize the witness operator by adjusting the reference state for asymmetrical mixed $W$ states,
% which usually requires higher efficiency to successfully detect entanglement.
%The proposed scheme seems to be feasible within the current technology considering particularly the experimental achievement 
%in Refs.\cite{Lombardi} and \cite{Pegg1}.

The author is very grateful to Otfried G{\"u}hne who alerted him with the notion of genuine multipartite entanglement 
along with the state in Eq.~(1) and also the improved witness operator in Ref.~\cite{Blatt}. 
He also acknowledges M. Suhail Zubairy for useful discussions. 
This work is supported by a grant from the Qatar National Research Fund. 

*email:hyunchul.nha@qatar.tamu.edu


\begin{references}
\bibitem{bi} A.~Peres, \prl {\bf 77}, 1413 (1996); M.~Horodecki, P.~Horodecki, and R.~Horodecki, Phys.~Lett.~A {\bf 223}, 1 (1996); 
W.~K.~Wootters, \prl {\bf 80}, 2245 (1998); R.~Simon, \prl {\bf 84}, 2726 (2000). 
\bibitem{Shor} P.~Shor, \pra {\bf 52}, R2493 (1995).
\bibitem{Cleve} R.~Cleve, D.~Gottesman, and H.-K.~Lo \prl {\bf 83}, 648 (1999).
\bibitem{onew} R.~Raussendorf and H.~J.~Briegel, \prl {\bf 86}, 5188 (2001).
\bibitem{multi} Z.-W.~Pan {\it et al.}, Nature (London) {\bf 403}, 515 (2000); 
C.~A.~Sackett {\it et al.}, {\it ibid} {\bf 404}, 256 (2000); O.~Mandel {\it et al.}, {\it ibid} {\bf 425}, 937 (2003); 
Z.~Zhao {\it et al.}, {\it ibid} {\bf 430}, 54 (2004); 
D.~Bouwmeester, J.-W.~Pan, M.~Daniell, H.~ Weinfurter, and A.~Zeilinger, \prl {\bf 82}, 1345 (1999); 
A.~Rauschenbeutel {\it et al.}, Science {\bf 288}, 2024 (2000); 
K.~J.~Resch, P.~Walther, and A.~Zeilinger, \prl {\bf 94}, 070402 (2005).
\bibitem{Guhne0} O.~G{\"u}hne, private communications, who remarked that the state in Eq.~(1) is known among some researchers 
in the community.
\bibitem{Acin} A.~Acin, D.~Bruss, M.~Lewenstein, and A.~Sanpera, \prl {\bf 87}, 040401 (2001).
\bibitem{Lewenstein} M.~Lewenstein, B.~Kraus, J.~I.~Cirac, and P.~Horodecki, \pra {\bf 62}, 052310 (2000).
\bibitem{Guhne1} O.~G{\"u}hne and P.~Hyllus, Int.~J.~Theor.~Phys. {\bf 42}, 1001 (2003).
\bibitem{Sanpera} M.~Bourennane {\it et al.},\prl {\bf 92}, 087902 (2004).
\bibitem{Toth} G.~Toth and O.~G{\"u}hne,\prl {\bf 94}, 060501 (2005).
\bibitem{Bell} Another possible method is to show much stronger violation of Bell inequalities for true $N$-qubit entangled states. 
N.~Gisin and B.~Bechmann-Pasquinucci, Phys.~Lett.~A {\bf 246}, 1 (1998); D.~Collins {\it et al.}, \pra {\bf 88}, 170405 (2002).
\bibitem{Eibl} The $W$ states involving polarization-entangled photons were experimentally demonstrated in M.~Eibl, N.~Kiesel, 
M.~Bourennane, C.~Kurtsiefer, and H.~Weinfurter, \prl {\bf 92}, 077901 (2004): H.~Mikami, Y.~Li, K.~Fukuoka, and T.~Kobayashi, 
\prl {\bf 95}, 150404 (2005): C.~F.~Roos {\it et al.}, Science {\bf 304}, 1478 (2004). 
\bibitem{nha0} H.~Nha and J.~Kim, \pra {\bf 75}, 012326 (2007).
\bibitem{nha1} H.~Nha and J.~Kim, \pra {\bf 74}, 012317 (2006).
\bibitem{nha2} H.~Nha, \pra {\bf 76}, 014305 (2007).
\bibitem{Hillery} M.~Hillery and M.~S.~Zubairy, \prl {\bf 96}, 050503 (2006). 
\bibitem{Agarwal} See also E.~Shchukin and W.~Vogel, \prl {\bf 95}, 230502 (2005);
G.~S.~Agarwal and A.~Biswas, New J.~Phys. {\bf 7}, 211 (2005). 
\bibitem{Dur} W.~Dur, G.~Vidal, and J.~I.~Cirac, \pra {\bf 62}, 62314 (2000).
\bibitem{Lombardi} Two-party single-photon entanglement and nonlocality were experimentally demonstrated. 
E.~Lombardi, F.~Sciarrino, S.~Popescu, and F.~De Martini, \prl {\bf 88}, 070402 (2002); B.~Hessmo, P.~Usachev, H.~Heydari, and 
G.~Bjork, \prl {\bf 92}, 180401 (2004).
\bibitem{Knill} E.~Knill, R.~Laflamme, R.~Martinez and C.-H.~Tseng, Nature {\bf 404}, 368 (2000); 
L.~M.~K.~Vandersypen and I.~L.~Chuang, Rev.~Mod.~Phys. {\bf 76}, 1037 (2005).
\bibitem{Lee} J.-S.~Lee and A.~K.~Khitrin, \apl {\bf 87}, 204109 (2005).\bibitem{Jozsa} R.~Jozsa, J.~Mod.~Opt. {\bf 41}, 2315 (1994); A.~Uhlmann, Re.~Math.~Phys. {\bf 9}, 273 (1976).  
\bibitem{Kim} M.~S.~Kim, J.~Lee, and W.~J.~Munro, \pra {\bf 66}, 030301 (2002).
\bibitem{nha3} See also H.~Nha and H.~J.~Carmichael, \pra {\bf 71}, 032336 (2005).
\bibitem{Campos} R.~A.~Campos, B.~E.~A.~Saleh, and M.~C.~Teich, \pra {\bf 40}, 1371 (1989).
\bibitem{Lvovsky} A.~I.~Lvovsky {\it et al.}, \prl {\bf 87}, 050402 (2001); 
A.~I.~Lvovsky and J.~Mlynek, {\it ibid.} {\bf 88}, 250401 (2002).
%\bibitem{Pegg1} K.~J.~Resch, J.~S.~Lundeen, and A.~M.~Steinberg, \prl {\bf 88}, 113601 (2002);
 S.~A.~Babichev, B.~Brezger, and A.~I.~Lvovsky, {\it ibid.} {\bf 92}, 047903 (2004). 
\bibitem{Blatt} H.~H{\"a}ffner {\it et al.}, Nature {\bf 438}, 643 (2005). 
\end{references}
\end{document}